\documentclass[12pt]{article}
\usepackage{epsfig}
\begin{document}
\newcommand{\bq}{\begin{equation}}
\newcommand{\eq}{\end{equation}}
\newcommand{\bqa}{\begin{eqnarray}}
\newcommand{\eqa}{\end{eqnarray}}
\newcommand{\nl}{\nonumber \\}
\newcommand{\al}{\alpha}
\newcommand{\be}{\beta}
\newcommand{\f}{\varphi}
\newcommand{\suml}{\sum\limits}
\newcommand{\la}{\lambda}
\newcommand{\ddv}[1]{{\partial\over\partial #1}}
\newcommand{\opi}{^{\mbox{{\tiny 1PI}}}}
\newcommand{\avg}[1]{\left\langle #1\right\rangle}
\newcommand{\dv}{\partial}
\newcommand{\eps}{\epsilon}
\newcommand{\plaat}[2]{\epsfig{figure=#1.eps,width=#2 cm}}

\begin{center}
{\bf {\Large Recursive actions for scalar theories
         \footnote{Supported by EU contract no. HPRN-CT-2000-00149}}}\\ 
\vspace*{2\baselineskip}
{\bf C.~Dams\footnote{{\tt chrisd@sci.kun.nl}}, 
R.~Kleiss\footnote{{\tt kleiss@sci.kun.nl}} }\\
University of Nijmegen, the Netherlands\\ 
\vspace*{1\baselineskip}
{\bf P.~Draggiotis\footnote{{\tt petros@sci.kun.nl}} },\\
University of Nijmegen, the Netherlands, and\\
NRCPS `Dimokritos', Agia Paraskevi, Athens, Greece\\ 
\vspace*{1\baselineskip}
{\bf E.~N.~Argyres, 
A.~van~Hameren\footnote{{\tt andrevh@sci.kun.nl}}, 
C.~G.~Papadopoulos\footnote{{\tt papadopo@alice.nuclear.demokritos.gr}} }\\
NRCPS `Dimokritos', Agia Paraskevi, Athens, Greece\\
 \vspace*{4\baselineskip}
Abstract
\end{center}
We introduce a class of self-interacting scalar theories in which the
various coupling contants obey a recursive relation. 
These imply a particularly simple form for the generating function
of the Feynman amplitudes with vanishing external momenta,
as well as for the effective potential.
In addition we discuss an interesting duality inherent in these models.
Specializing to the case of zero spacetime dimensions we find
intriguing nullification properties for the amplitudes.

\section{Introduction}

In this paper we discuss a special class of Euclidean 
theories of self-interacting
scalar fields. More in particular, we study amplitudes with
vanishing external momentum, as for instance implied in the definition
of an effective potential; but also the special case of theories in
zero spacetime dimensions is subsumed (the zero-dimensional case is
of course of paradigmatic interest because the path integral is 
a simple integral, amenable to straightforward solution and manipulation).
Most of our results will, in fact, be derived for zero dimensions.
Zero-dimensional field theories have been amply discussed: apart from the
useful introductory treatment in \cite{cvit} we may refer to
\cite{goldberg,thooft,bender,zero} as recent applications. The aim of this paper
is to study properties of essentially non-polynomial theories in which the
coupling constants obey a simple algebraic relation. In section 2, 
we define {\em recursive\/} theories, and show how the various zero-momentum
Green's functions are related to one another in a surprisingly simple manner,
which allows us to express the complete set $n$-particle amplitudes in terms
of the tadpole alone. In section 3, we study a duality inherent in our models,
that relates the elementary field in one theory with certain composite field
in its dual. In section 4 we discuss several explicitly solvable
zero-dimensional recursive theories. In section 5, we tackle the structure
of loop corrections for general zero-dimensional self-interacting scalar
theories, with special emphasis on the occurrence of `nullification', that
is, a special choice of parameters for which {\em all\/} loop corrections
of a given loop order vanish. Section 6 is devoted to nullification
in recursive thories, where an intriguing pattern is exhibited. In section 7,
we address the application of renormalization in zero dimensions, in the
spirit of \cite{cvit}.

\section{Feynman amplitudes for recursive actions}
We consider self-interacting theories of a field $\f$ with mass $m$
in $d$ Euclidean spacetime dimensions, with potentials given by
\bq
V(\f) = \suml_{n\ge3}{\la_n\over n!}\f^n\;\;.
\label{potentialdefintion}
\eq
We also introduce $\la_2\equiv\mu = m^2$. Note that in the sum $n$ runs,
in principle, all the way up to infinity.
Our class of models is characterized by the following property:
there exist (dimensionful) constants $\al$ and $\be$ such that
\bq
\la_{n+1} = \la_n(\al n + \be)\;\;,\;\;n\ge2\;\;.
\eq
Of course, we can determine $\al,\be$ from
\bq
\al = {\la_4\over\la_3}-{\la_3\over\mu}\;\;\;,\;\;\;
\be = 3{\la_3\over\mu} - 2{\la_4\over\la_3}\;\;,
\eq
and the combination $\al\f$ is dimensionless.
The Lagrangian density of these models is given by
\bq
{\cal L} = {1\over2}(\vec{\nabla}\f)^2 + 
{\mu\over p\be}\left((1-\al\f)^{-\be/\al}-1-\be\f\right)\;\;,
\;\;p=\al+\be\;\;,
\eq
and in zero dimensions the action itself is simply
\bq
S(\f) ={\mu\over p\be}\left((1-\al\f)^{-\be/\al}-1-\be\f\right)\;\;.
\eq
We call these models {\em recursive\/}.
When $-\be/\al$ is an integer $K\ge2$, the potential is a finite polynomial
of order $K$: $\f^3$ theory is recursive, but neither pure $\f^4$ theory
nor spontaneously broken $\f^4$ theory are recursive.\\

Consider a connected Feynman diagram $D(n)$,
entering into the $1\to n$ amplitude.
Let this graph have $I$ internal lines, $E=1+n$ external lines,
$V_k$ vertices of type $\f^k$ and $L$ loops. We have
\bq
D(n) \propto\; \hbar^L\mu^{-I-E+dL/2}\la_3^{V_3}\la_4^{V_4}\cdots\;\;.
\eq
Here we have assumed that the regularization of the loop integrals
is performed in a manner that does not introduce another physical mass scale
(as would be the case in say, Pauli-Villars regularization), so that
a $d$-dimensional loop integral contributes a factor $m^d=\mu^{d/2}$;
an example is dimensional regularization. In that case there enters,
of course, an `engineering dimension', which we include in the
(possibly very complicated) proportionality constant.
Note that this is consistent also for $d=0$ since then loop integrals
are simply absent.

The two topological relations
\bq
\suml_{k\ge3}kV_k = 2I+E\;\;\;,\;\;\;
\suml_{k\ge3}V_k = I+1-L\;\;,
\eq
can be written as follows for this diagram:
\bqa
\suml_{k\ge2}k\la_k\ddv{\la_k}D(n) &=&
d\hbar\ddv{\hbar}D(n) - (n+1)D(n)\;\;,\nl
\suml_{k\ge2}\la_k\ddv{\la_k}D(n) &=&
({d\over2}-1)\hbar\ddv{\hbar}D(n) - nD(n)\;\;.
\eqa
Since this holds for any $D(n)$, it holds {\it a fortiori\/} also for the
full $1\to n$ amplitude $a(n)$.

Let us now consider what happens if we add one external leg to the
amplitudes. This may be done in several ways. In the first place, we may
simply attach the external line to any $\f^k$ vertex, thereby turning it into 
a $\f^{k+1}$ vertex, giving a factor $V_k\la_{k+1}/(\la_k\mu)$.
In other words, attaching a line to any vertex in the diagram
is equivalent to the operation
$$
D(n) \to \suml_{k\ge3}{\la_{k+1}\over\mu}\ddv{\la_k}D(n)\;\;\;.
$$
In the second place,
we may attach the external line to any line by a three-point vertex. If the
momentum flowing in the original line is $q$, the attachment
turns $(q^2+\mu)^{-1}$  into $-\la_3\mu^{-1}(q^2+\mu)^{-2}$, so that
we can write this procedure as
$$
D(n) \to {\la_3\over\mu}\ddv{\mu}D(n)\;\;\;.
$$
Note that this also works for internal lines in
loop diagrams, owing to the fact that the external momenta all vanish. 
In this way, we can form all amplitudes
from the vacuum bubbles of the theory, with the single exception of the
bare propagator. We also want to stress that, in this procedure,
the symmetry factors of all diagrams will come out
correct automatically: in a sense, our procedure is how the symmetry
factors are defined in the first place anyway.
We therefore have the following recursion
between amplitudes:
\bq
\mu\;a(n+1) = \delta_{n,0} + \suml_{k\ge2}\la_{k+1}\ddv{\la_k}a(n)
\;\;\;,\;\;n\ge0\;\;.
\eq
So far this is general for zero-momentum amplitudes. In the case of recursive
actions, we can use the relation between $\la_{k+1}$ and $\la_k$ to good effect.
Let us denote by $\phi(x)$ the generating function of all $1\to n$ amplitudes:
\bq
\phi(x) = \suml_{n\ge0}a(n){x^n\over n!}\;\;.
\eq
For recursive actions, this then satisfies the differential equation
\bq
(\mu+px)\ddv{x}\phi + \al\phi + \left((1-d/2)\be-d\al\right)\hbar\ddv{\hbar}\phi = 1\;\;.
\eq
The solution can be written as
\bqa
\phi(x) &=& {1\over\al}\left(1-{1\over(1+px/\mu)^{\al/p}}\right)\nl
&+& {1\over(1+px/\mu)^{\al/p}}
\suml_{L\ge1}t_L(\mu;\al,\be)
\left({\hbar\over(1+px/\mu)^{(\be(1-d/2)-d\al)/p}}\right)^L\;\;.
\eqa
we see that {\em for recursive actions, all connected amplitudes are
completely determined by the tadpole} $\phi(0)=\suml_Lt_L\hbar^L$.
The above argument does not, however, allow us to
determine the tadpole itself.
This reflects the fact that, whereas the operation of adding
an extra external {\em line\/} is,  in the above, formulated as a fairly simple
algorithm, the operation of adding an extra {\em loop\/} does not appear to
follow any simple algorithm yielding the right symmetry factors.

In zero dimensions, simple dimensional analysis shows that
$t_L(\mu;\al,\be)$ can be written as $\mu^{-L}$ times an expression
in $\al$ and $\be$
that is homogeneous of degree $2L-1$.
Moreover, since the tadpole contains $\al$ and $\be$ only through
the coupling constants $\la_k$, which only enter in the numerator
of any Feynman diagram, we conclude that, for $p\ne0$,
\bq
t_L = {p^{2L-1}\over\mu^L}R_{2L-1}(u)\;\;,
\eq
where $R_q$ is a polynomial of degree $q$ and the ratio between $\al$
and $\be$ is encoded in
\bq
u = {2\al+\be\over\al+\be} = 1 + \al/p\;\;.
\eq
It is clear that we may put $\mu=p=1$ without loss of generality (except
for the special case $\al+\be=0$), since they can always be put back into
any expression.

Finally, by judicious choice of $\al$ and $\be$ we can single out theories
with interesting properties. For instance, 
when $\al/p$ is a negative integer $-n$, the tree-level
amplitude generating function $\phi_0(x)$ is a finite polynomial
of degree $n$. This implies that the tree level amplitudes with $n+2$ or more
external legs all vanish. In higher orders, though, these amplitudes
may be nonzero since $\phi_L(x)$ has exponent $n-(n+1)L$.
In the same vein, when $p=0$ the amplitudes do not necessarily vanish,
but the generating function goes with a power of $\exp(-x)$ at every loop
order, which implies that the amplitudes go as $1/n!$ rather than as
$n!$, a  situation that has been discussed (at the kinematic threshold rather
than at zero momentum) elsewhere \cite{unit}.

Similarly,
if $\be/\al=-n/(n+1)$, with $n$
a nonnegative integer, the $L$-loop contribution $\phi_L(x)$ is
a finite polynnomial in $x$ of degree $n(L-1)-1$ for $L\ge2$, which implies
that $L$-loop corrections vanish for Green's functions with more that
$n(L-1)$ external legs.\\

In the one-dimensional case, an interesting situation is that where
$\be=2\al$. In that case,
\bq
\phi(x) = {1\over\al}-{1\over(1+px/\mu)^{\al/p}}
\left({1\over\al}-\suml_{L\ge1}\hbar^Lt_L\right)\;\;,
\eq
so that in the zero-momentum amplitudes 
{\em all\/} loop corrections can be completely absorbed into
finite tadpole and mass renormalization (see also the discussion
in a later section of this paper).

For theories with $\be/\al=-2n/(1+2n)$ we find, in this case,
that $\phi_L(x)$ contains the exponent $(1+3n)(L-1)+n$, so that again
loop corrections vanish for sufficiently large numbers of legs.\\

In higher dimensions, the tadpole factors $t_L$ will themselves also
depend on $d$, and in fact for $d\ge2$ they contain divergences.
As long as we use dimensional regularization it is therefore tempting, but
erroneous, to choose attractive-looking values for $d$. For instance,
choosing $d=2$ and $\al=0$ would at first sight seem to eliminate
the $x$ dependence in the loop corrections, but the more careful treatment
$d=2-2\eps$, $\eps\to0$ reveals that $t_L$ will, in general contain
poles in $\eps$ up to $\eps^{-L}$ so that, in fact all amplitudes
have loop corrections.\\

We may use a similar argument for the
one-particle irreducible (1PI) diagrams of the theory.
The only difference is that the new line may {\em not\/} be attached to
an existing external line, so that the relevant operation reads
$$
D(n) \to \suml_{k\ge2}{\la_{k+1}\over\mu}\ddv{\la_k}D(n)
 + (n+1){\la_3\over\mu^2}D(n)\;\;\;.
$$
Denoting the generating function of the 1PI $1\to n$
amplitudes by $\phi\opi$, we now find the differential equation
\bq
(\mu-\al x)\ddv{x}\phi\opi - p\phi\opi + 
\left((1-d/2)\be-d\al\right)\hbar\ddv{\hbar}\phi\opi = 1\;\;,
\eq
which is quite similar to the equation for the connected amplitudes.
Its solution reads
\bqa
\phi\opi(x) &=& -{1\over p} + {1\over p}(1-\al x/\mu)^{-p/\al} \nl
&+& (1-\al x/\mu)^{-p/\al}\suml_{L\ge1}
t_L\opi(\mu;\al,\be)
\left(\hbar(1-\al x/\mu)^{(\be(1-d/2)-d\al)/\al}\right)^L\;\;,
\eqa
and again everything is determined by the (1PI) tadpole. The effective action
of the theory, $\Gamma(\f)$, can simply be found from
\bq
\Gamma'(\f) = \mu\phi\opi(\mu\f)\;\;.
\eq

A final note on the divergence structure of the amplitudes is in order. 
On first 
sight it might be thought that, since any $L$-loop amplitude is proportional
to $t_L$, these amplitudes must all have the same divergence structure.
That this is not necessarily the case can be seen from the following simple
example. Consider the 1PI one-loop amplitudes for the pure $\f^3$ theory in $d$
dimensions, {\it i.e.\/} $\be=-3\al$ and $L=1$. We find
\bqa
\phi\opi_1(x) &=& \hbar t_1 \left(1-{\al x\over\mu}\right)^{d/2-1}\nl
&=& \hbar t_1\left(1 - \left({\al x\over\mu}\right){(d-2)\over2}
+ \left({\al x\over\mu}\right)^2{(d-2)(d-4)\over8} + \cdots\right)\;\;.
\eqa
The tadpole contribution,
\bq
\hbar t_1 =
 -{\al\hbar\over2(4\pi)^{d/2}}\mu^{d/2-1}\Gamma\left(1-{d\over2}\right)\;\;,
\eq
is divergent for $d=2,4,6,\ldots$. It can be seen that in two dimensions
the 1PI propagator is then finite. In four dimensions, the 1PI propagator is
divergent (but less so than the tadpole) and the 1PI three-point function
is finite, and so on; precisely in accordance with what is expected on the
basis of the Feyman diagrams, that all consist of a single loop beaded
with three-point vertices.\\

Before finishing this section we wish to point out the following. In the 
definition of the recursive action it is very important that the one-point
coupling $\la_1$ is {\em not\/} included. Consider a theory
 with an explicit tadpole term, given by
\bq
V_{\mbox{tad}}(\f) = \la_1\f + V(\f)\;\;,
\eq
with a recursive potential $V(\f)$ as given in 
Eq.(\ref{potentialdefintion}).
It is easily seen that the generating function $\phi_{\mbox{tad}}(x)$ of this
theory is related to the one without the tadpole term as
\bq
\phi_{\mbox{tad}}(x) = \phi(x-\la_1) =
{1\over\al} - {1\over\al\left(1-{\la_1p\over\mu}+{xp\over\mu}\right)^{\al/p}}
+ \mbox{(loop corrections)}\;\;.
\eq
If we let $\la_1$ approach its recursive value, $\la_1=\mu/p$, the generating
function obtains its singularity at $x=0$, and  perturbation theory breaks
down even at the tree level. This is caused by the fact that, with a nonzero
bare tadpole term, every amplitude at every loop order contains an infinite
number of Feynman diagrams, the sum of which, order
by order, is no longer convergent at
precisely the recursive value of the tadpole. For larger absolute values
of the tadpole coupling, $\phi$ can only be obtained by analytic continuation.

\section{Duality in zero dimensions}
Throughout this section, we shall assume $d=0$. The Euclidean
path integral in the presence of a source $x$ is then given by
\bq
Z(x) = \int\limits_C\;d\f\;\exp\left(-{1\over\hbar}(S_{\al,\be}(\f)-x\f)\right)\;\;,
\eq
where we have explicitly indicated the parameters entering in $S(\f)$. 
The integration contour $C$ is preferably such that the integrand
vanishes sufficicently fast at the endpoints. However, if we restrict ourselves
to perturbation theory it is sufficient that the endpoints do not
approach the perturbative extremum $\f=0$. This is particularly important
when $\be/\al$ is not integer so that the action displays branch cuts
starting at $\f=1/\al$. The generating funciton of the connected 
amplitudes is given by
\bq
\phi(x) = {\hbar\over Z(x)}\ddv{x}Z(x)\;\;.
\eq
The tadpole for this theory is therefore
\bq
T_{\al,\be} \equiv \phi(0) = {\avg{\f}_{\al,\be}\over\avg{1}_{\al,\be}}\;\;,
\eq
where
\bq
\avg{A(\f)}_{\al,\be} \equiv \int\;d\f\;A(\f)
\exp\left(-{S_{\al,\be}(\f)\over\hbar}
\right)
\eq
for any function $A(\f)$ of $\f$. By partial integration we may prove the
following useful lemma:
\bq
\avg{A(\f)S'_{\al,\be}(\f)}_{\al,\be} = \hbar\avg{A'(\f)}_{\al,\be}\;\;.
\eq
Now, consider the object $\psi$ dual to $\f$, defined by
\bq
(1-\be\psi)^\al(1-\al\f)^\be = 1\;\;\;\Rightarrow\;\;\;
\psi = {1\over\be}\left(1-(1-\al\f)^{-\be/\al}\right)\;\;.
\eq
In terms of $\f$, $\psi$ is a `composite' object, and we have
\bq
S_{\al,\be}(\f) = -{\mu\over p}(\f+\psi) = S_{\be,\al}(\psi)\;\;.
\eq
We may replace $\f$ as a dummy integration variable by $\psi$.
For instance,
\bq
\avg{1}_{\al,\be}= \avg{{d\f\over d\psi}}_{\be,\al} =
 \avg{-1-{p\over\mu}S'_{\be,\al}(\psi)}_{\be,\al}
= -\avg{1}_{\be,\al}\;\;.
\eq
The zero-dimensional action satisfies a linear differential equation:
\bq
(1-\al\f)S'_{\al,\be}(\f) - \be S_{\al,\be}(\f) = \mu\f\;\;,
\eq
so that $S$ can be expressed in terms of $S'$.
Together with the lemma, this allows us to compute also
\bqa
\al\avg{\f}_{\al,\be} &=&
\al\avg{\left(1+{p\over\mu}S'_{\be,\al}(\psi)\right)
\left(\psi+{p\over\mu}S_{\be,\al}(\psi)\right)}_{\be,\al}\nl &=&
{p\hbar\over\mu}(\al-\be)\avg{1}_{\be,\al}
-\be\avg{\psi}_{\be,\al}\;\;.
\eqa
We therefore find a simple relation between the tadpoles of the two dual
theories with $S_{\al,\be}$ and $S_{\be,\al}$:
\bq
\al\left(T_{\al,\be}(\hbar)+{p\hbar\over\mu}\right) =
\be\left(T_{\be,\al}(\hbar)+{p\hbar\over\mu}\right)\;\;.
\eq
This duality allows for some immediate conclusions. In the first place,
the free theory is recursive, with $\be=-2\al$ so that $\la_3=0$, and it has
a vanishing tadpole. Its dual is the action given by
\bq
S_{-2\al,\al}(\f) = {\mu\over\al^2}\left(
1+\al\f - \sqrt{1+2\al\f}\right)\;\;,
\eq
and its tadpole is therefore immediately seen to be
\bq
T_{-2\al,\al} = {3\al\hbar\over2\mu}\;\;; \label{squarerootactiontadpole}
\eq
in this theory, {\em all\/} loop corrections beyond the one-loop level
vanish identically!
Similarly, the action with $\be\to0$ and hence $u=2$, given by
\bq
S_{\al,0}(\f) = {\mu\over\al^2}\left(
-\al\f - \log(1-\al\f)\right)\;\;,
\eq
has for its tadpole
\bq
T_{\al,0} = -{\al\hbar\over\mu}\;\;,  \label{logactointadpole}
\eq
and again all higher orders vanish identically. The results
(\ref{squarerootactiontadpole}) and (\ref{logactointadpole}) are 
confirmed by explicit computation of the zero-dimensional path integral.\\

The zero-dimensional duality has another interesting 
con\-se\-quen\-ce. Put\-ting
$\mu=p=1$ and writing $\al=u-1$, $\be=2-u$ we can write the tadpole
du\-a\-li\-ty as
\bq
(u-1)R_{2L-1}(u) = (2-u)R_{2L-1}(3-u)\;\;,\;\;L\ge2\;\;,
\eq
where $R_{2L-1}(u)$ is the polynomial entering in $t_L$
as discussed above. This means that $R_{2L-1}(u)$ must have a root at
$u=2$ (as indeed we have seen). Moreover, since $u=0$ corresponds to the
free action, $R_{2L-1}(u)$ must vanish for $u=0$, and by duality
also for $u=3$. We can therefore
write
\bq
R_{2L-1}(u) = u(2-u)(3-u)P_{L-2}(\omega)\;\;\;,\;\;\;
\omega = u(3-u)\;\;\;,\;\;\;L\ge2,
\eq
where $P_{L-2}(\omega)$ is a polynomial of degree $L-2$ only.

\section{Explicit solutions in zero dimensions}
In this section we discuss a few explicitly solvable models with $d=0$.
For simplicity, we shall take $\mu=0$ and $p=1$. The models are, then,
completely specified by the parameter $u$.
We may therefore write $S_{\al,\be}(\f)=S_u(\f)$,
and the duality operation is the interchange $u\leftrightarrow3-u$.
The partition function $Z(x)$ is in general determined from 
the Schwinger-Dyson (SD) equation
\bq
S'\left(\hbar\ddv{x}\right)Z(x) = xZ(x)\;\;,
\eq
which leaves the overall normalization of $Z(x)$ undetermined.
For recursive actions, the SD equation may be rewritten as
\bq
\left(1-\al\hbar\ddv{x}\right)^{-1/\al}Z(x) = (1+x)Z(x)\;\;.
\eq

The simplest case is the free theory, $u=0$, leading to $Z(x)=\exp(x^2/2)$
and $\phi=x$; and the effective action is $\Gamma_0(\f)=S_0(\f)$.
The next simplest case is $u=2$:
\bq
S_2(\f) = -\f - \log(1-\f)\;\;,
\eq
leading to the SD equation
\bq
{1\over1-\hbar\dv}Z(x) = (1+x)Z(x)\;\;\;,\;\;\;\dv\equiv\ddv{x}\;\;.
\eq
Multiplying from the left by $Z(x)^{-1}(1-\hbar\dv)$ on both sides
gives immediately the form of $\phi$:
\bq
\phi = {x-\hbar\over(1+x)}\;\;,
\eq
in agreement with the result from duality. The effective action is most simply
obtained from inverting this relation:
$$
\phi(x) = F(x) \to x(\phi) = \Gamma'(\phi)\;\;.
$$
In this case, we find
\bq
\Gamma_2(\f) = -\f - (1+\hbar)\log(1-\f)\;\;,
\eq
so that also the effective action is free of $L\ge2$ corrections.\\

The case $u=1$ corresponds to the action dual to $S_2(\f)$:
\bq
S_1(\f) = e^\f - 1 - \f\;\;,
\eq
leading to a {\em functional\/} form for the SD equation:
\bq
e^{\hbar\dv}Z(x) = Z(x+\hbar) = (1+x)Z(x)\;\;,
\eq
and a functional equation for $\phi(x)$:
\bq
\phi(x+\hbar) = \phi(x) + {\hbar\over1+x}\;\;.
\eq
Together with the requirement $\lim_{\hbar\to0}\phi(0)=0$ this
implies
\bqa
\phi(x) &=& \log\hbar + \psi\left({1+x)\over\hbar}\right)\nl
&=& \log(1+x) - {\hbar\over2(1+x)}
-\suml_{L\ge2}{B_{L}\over L}\left({\hbar\over1+x}\right)^L\;\;,
\eqa
where $\psi()$ denotes the digamma function, 
and we have indicated the asymptotic expansion, where $B_L$ are the
Bernoulli numbers. 
The behaviour with $x$ of this result is, of course,
already given from the recursivity of the model, but as stated
before the tadpole itself can only be obtained from
the SD equation.
Since $B_L=0$ for odd $L\ge3$, we conclude that
for this model {\em all odd-loop amplitudes beyond the one-loop level
vanish completely}. This result is significant since, in this model,
every coupling constant $\la_n$ is unity, and the value of every Feynman
diagram is given by only its symmetry factor times a factor $(-1)$ for
every vertex; we therefore have, here, a strictly graph-theoretic result.
The one-loop tadpole consists, of course, of only a single diagram and can
never vanish.
For the effective action we find a similar result, at least empirically.
Writing $x$ as a function of $\phi$ gives the effective action:
\bq
\Gamma_1'(\f) = -1 + e^\f + \suml_{L\ge1}a_L\hbar^Le^{(1-L)\f}\;\;,
\eq
where $a_1=1/2$ and the first few even coefficients $a_L$ read
\bqa
&&a_2 = -{1/24}\;\;,\nl
&&a_4 = {3/640}\;\;,\nl
&&a_6 = -{1525/580608}\;\;,\nl
&&a_8 = {615881/199065600}\;\;,\nl
&&a_{10}=-{3058641/504627200}\;\;,\nl
&&a_{12} ={38800188510523/2191186722816000}\;\;,\nl
&&a_{14}=-{3213747182969063/44497945755648000}\;\;,\nl
&&a_{16}={100462329712125/255806104666112}\;\;.
\eqa
The only one-loop 1PI amplitude is the tadpole.
All coefficients for odd $L\ge3$ appear to vanish again: we have checked this
up to 60 loops, but we have not been able
to prove it rigorously. The fact that the pattern of zeroes in both $\phi$ and
$\phi\opi$ is the same is intimately tied up with the occurrence of the
Bernoulli numbers: if we assume the given $x$ dependence in $\phi$ and
insist that the 3,5,7,$\ldots$ loop corrections vanish in both $\phi$ and
$\phi\opi$ we recover the above result for $\phi$ as the unique solution.\\

For $u=3$ we have the action dual to the free one:
\bq
S_3(\f) = 1 - \f - \sqrt{1-2\f}\;\;.
\eq
Its SD equation reads
\bq
DZ = yZ\;\;\;,\;\;\;y=1+x\;\;\;,\;\;\;D\equiv (1 - 2\hbar\ddv{y})^{-1/2}\;\;.
\eq
Although this is an infinite-order differential equation,
it is solvable by using the fact that $D$ and $y$ obey a commutation relation
\bq
[D,y] = \hbar D^3\;\;.
\eq
This allows us to write
\bq
D^2Z = D(yZ) = yDZ + \hbar D^3Z = y^2Z + \hbar D^2(yZ)\;\;.
\eq
Multiplying from the left by $D^{-2}=1 - 2\hbar\dv/\dv y$ then gives
a linear SD equation for $Z$, from which $\phi$ follows algebraically:
\bq
Z = y^2Z - 3\hbar yZ - 2\hbar y^2\dv Z\;\;\;,\;\;\;
\phi = {1\over2}\left(1-{1\over(1+x)^2}\right)-{3\hbar\over2}{1\over(1+x)}\;\;,
\eq
which is precisely the result obtained earlier by duality.
The effective action is given by
\bq
\Gamma_3'(\f) = -1 + {1\over1-2\f}\left[
{3\hbar\over2} + \sqrt{1-2\f+\left({3\hbar\over2}\right)^2}
\right]\;\;.
\eq
For this theory, the 1PI amplitudes for odd $L\ge3$ vanish identically.
\\

For polynomial recursive actions with highest interaction term $\f^{K+1}$
(integer $K$) and $p=\mu=1$ we have $\al=-1/K$, $\be=(K+1)/K$, and hence the
SD equation has finite order:
\bq
\left(1+{\hbar\over K}\ddv{y}\right)^KZ = yZ\;\;.
\eq
We can show that, even though the dual theories of these actions are not
of finite polynomial form, nevertheless their SD equations can also be cast in
the form of differential equations of order $K$, as follows.
The dual actions have $\al=(K+1)/K$, and the SD equation is
\bq
\left(1-{\hbar(K+1)\over K}\ddv{y}\right)^{-K/(K+1)}Z = D^KZ = yZ\;\;,
\eq
where
\bq
D = \left(1-{\hbar(K+1)\over K}\ddv{y}\right)^{-1/(K+1)}\;\;.
\eq
The differential operator $D$ has the following commutation relation with
$y$:
\bq
D^sy = yD^s + {s\hbar\over K}D^{s+K+1}\;\;\;,
\eq\
for general $s$, and hence
\bq
D^sZ = D^{s-K}(yZ) = yD^{s-K}Z + {\hbar(s-K)\over K}D^{s+1}Z\;\;.
\eq
By repeating this operation, it is easily seen that
\bqa
A_0 & \equiv & D^sZ = A_1 = A_2 = A_3 =\ldots,\nl
A_m &=& \suml_{n=0}^m \gamma_n^{(m)} y^{m-n}\hbar^n
D^{s-mK+n(K+1)}Z\;\;,\nl
\gamma_n^{(m)} &=& \gamma_n^{(m-1)} +
\gamma_{n-1}^{(m-1)}{s-(m+1)K+n(K+1)-1\over K}\;\;,
\eqa
and the recursion relation for the $\gamma$'s starts at 
$\gamma_n^{(0)}=\delta_{n,0}$. Choosing $s=K(K+1)$ and $m=K$
we are then left with
\bq
D^{K(K+1)}Z = \suml_{n=0}^K 
\gamma_n^{(K)} y^{K-n}\hbar^n D^{n(K+1)+K}Z\;\;,
\eq
or, in other words,
\bq
\left(1-{\hbar(K+1)\over K}\ddv{y}\right)^{-K}\!\!\!Z
= \suml_{n=0}^K \gamma_n^{(K)} y^{K-n}\hbar^n
\left(1-{\hbar(K+1)\over K}\ddv{y}\right)^{-n}\!\!\!(yZ)\;\;.
\eq
By multiplying from the left by $\left(1-{\hbar(K+1)\over K}\ddv{y}\right)^{K}$,
we obtain the differential equation of finite order $K$ mentioned above,
with coefficients containing powers of $y$ up to $y^{K+1}$. We give here
the results for the first few $K$ values:
\bqa
K=1 &:& Z_0 = y^2Z_1 - 3\hbar yZ_0\;\;,\nl
K=2 &;& Z_0 = y^3Z_2 - 6\hbar y^2Z_1 + 5\hbar^2yZ_0\;\;,\nl
K=3 &:& Z_0 = y^4Z_3 - 10\hbar y^3Z_2 + {65\over3}\hbar^2y^2Z_1
-{70\over9}\hbar^3yZ_0\;\;,\nl
K=4 &:& Z_0 = y^5Z_4 - 15\hbar y^4Z_3 + 60\hbar^2y^3Z_2\nl 
&&\hphantom{ Z_0 = }
-{525\over8}\hbar^3y^2Z_1 + {189\over16}\hbar^4yZ_0\;\;,
\eqa
where
$Z_n \equiv \left(1-{\hbar(K+1)\over K}\ddv{y}\right)^nZ$.\\

Another class of models is that for which $\al=1/K$, with $K$ a positive
integer, hence $u=1+1/K$. Their SD equation reads
\bq
Z = \left(1-{\hbar\over K}\ddv{y}\right)^K(yZ)\;\;,
\eq
again a linear equation of finite order. Among these models is the
`self-dual' action with $K=2$ and $u=3/2$, with solution
\bq
Z = y^{-1/2}\exp\left(-{2y\over\hbar}\right)I_1\left({4y^{1/2}\over\hbar}\right)
\;\;,
\eq
where $I$ is the modified Bessel function of the first kind. The resulting
tadpole reads
\bq
T(\hbar) = 2{I_0\left(4/\hbar\right)\over 
I_1\left(4/\hbar\right)}-2-\hbar\;\;.
\eq
The other solution, which has the modified Bessel function of the second kind,
$K_1$ instead of $I_1$ in $Z(x)$, has a nonvanishing tree-level tadpole
and hence does not correspond to a perturbative solution.
For $K\to\infty$ we return to the case $u=1$ discussed above.\\

A final case of interest is the `almost-free' theory, with
$u=\eps\ll1$. The action reads, in this case
\bq
S_\eps(\f) = {1\over2}\f^2 + {\eps\over4}\left(
-2\f-3\f^2 + 2(1+\f)^2\log(1+\f)\right) + {\cal O}(\eps^2)\;\;,
\eq
and the SD equation,
\bq
\left(
(1-\eps)\hbar\ddv{x}+
\eps\left(1+\hbar\ddv{x}\right)\log\left(1+\hbar\ddv{x}\right) -x \right)
Z(x) = {\cal O}(\eps^2)\;\;,
\eq
looks quite hopeless. However, we may solve it by realizing that, for
small $\eps$, the coupling constants are also very small:
\bq
\lambda_k = (-)^{(k-3)}(k-3)!\eps + {\cal O}(\eps^2)\;\;,\;\;k\ge3\;\;.
\eq
Therefore, the tadpole is dominated by diagrams with only one vertex. 
The $L$-loop tadpole therefore contains only $\lambda_{2L+1}$, and
\bq
T(\hbar) = -\suml_{L\ge1}{(2L-2)!\over 2^L L!}\hbar^L\eps =
-{\eps\over2}\int\limits_0^\infty dz\;{\exp(-z)\over z}\left(
1-\sqrt{1-2z\hbar}\right)\;\;.
\eq
In the integral representation, the ambiguity arising from the branch cut
shows up as a nonpertrubative effect only.
Note that this result allows us to conclude that
\bq
P_{L-2}(0) = -{(2L-2)!\over 2^{L+1}\cdot3\cdot L!}\;\;,\;\;L\ge2\;\;.
\eq

\section{Higher loops and nullification patterns}
The observed patterns of vanishing higher loop corrections leads naturally
to the question of whether there are more such patterns, maybe for
non-recursive actions. To answer this it is necessary to determine the
general structure of higher-loop corrections in general zero-dimensional
theories. To this end, we may write the $L$-loop term in the SD equation as
\bq
\suml_{\{c_{l,n}\}}S^{(k)}\left(\phi_0(x)\right)\prod\limits_{l,n\ge0}
\left[{1\over c_{l,n}!}
\left(\phi_l^{(n)}(x){\hbar^n\over(n+1)!}\right)^{c_{l,n}}\right]
=0\;\;,\label{biggy}
\eq
where the sum extends over all nonnegative integer values of $c_{l,n}$ with the
condition that $\sum(l+n)c_{l,n}=L$, and $k=1+\sum(n+1)c_{l,n}$: the upper
indices in brackets denote derivatives. 
This equation is valid for general zero-dimensional actions: all specifics of
the action are encoded in $\phi_0(x)$, or rather $f(x)=\phi'_0(x)$: 
since by definition $S'(\phi_0(x))=x$, we can find the higher derivative terms
using
\bq
S^{(k)}(\phi_0(x)) = {1\over f(x)}\ddv{x}S^{(k-1)}(\phi_0(x))\;\;.
\eq
In the above representation of the SD equation, the $L$-loop correction
$\phi_L(x)$ occurs only in the combination $\phi_L(x)S^{(2)}(\phi_0(x))$, and
therefore $\phi_L(x)$ is simply expressed in terms of the lower ones and their
derivatives, and hence eventually in terms of $f(x)$ and its derivatives.
The first few loop corrections are
\bqa
\phi_1 &=& {1\over2}f_1\;\;,\nl
\phi_2 &=& {1\over24f}\left(12f_{1}^3-14f_{2}f_{1}+3f_{3}\right)\;\;,\nl
\phi_3 &=& {-1\over48f^2}\left(-144f_{1}^5-68f_{1}^2f_{3}+11f_{4}f_{1}-f_{5}-96f_{1}f_{2}^2
\right.\nl &&\left.+276f_{1}^3f_{2}+20f_{3}f_{2}\right)\;\;,\nl
\phi_4 &=& {1\over5760f^3}\left(138480f_{1}^4f_{3}-1212f_{4}f_{3}-136800f_{1}^2f_{2}f_{3}+13260f_{2}^2f_{3}
\right.\nl &&\left.
+9360f_{1}f_{3}^2+204480f_{1}^7-300f_{6}f_{1}+15f_{7}-780f_{5}f_{2}+3320f_{5}f_{1}^2
\right. \nl && \left.
+390960f_{1}^3f_{2}^2-545280f_{1}^5f_{2}+14620f_{1}f_{4}f_{2}
\right.\nl &&\left.-25200f_{1}^3f_{4}-64440f_{1}f_{2}^3\right)\;\;,\nl
\phi_5 &=& {1\over11520f^4}\left(8400f_{3}^3+3f_{9}+7292160f_{1}^9-152520f_{1}^2f_{5}f_{2}+9760f_{6}f_{1}f_{2}
\right. \nl && \left.
+17872f_{1}f_{5}f_{3}+1552320f_{1}^3f_{2}f_{4}-237560f_{1}^2f_{3}f_{4}+39400f_{2}f_{3}f_{4}
\right. \nl && \left.
-335820f_{1}f_{2}^2f_{4}-430000f_{1}f_{2}f_{3}^2+4212960f_{1}^2f_{2}^2f_{3}-11004960f_{1}^4f_{2}f_{3}
\right. \nl && \left.
-202800f_{2}^3f_{3}-320f_{7}f_{2}-19640f_{1}^3f_{6}+1640f_{7}f_{1}^2+992160f_{2}^4f_{1}
\right. \nl && \left.
-10357920f_{1}^3f_{2}^3-95f_{8}f_{1}+27145440f_{1}^5f_{2}^2-24914880f_{1}^7f_{2}
\right. \nl && \left.
+994240f_{1}^3f_{3}^2+6411840f_{1}^6f_{3}-962f_{5}f_{4}-672f_{6}f_{3}+12660f_{2}^2f_{5}
\right. \nl && \left.
-1217040f_{1}^5f_{4}+10870f_{1}f_{4}^2+176400f_{1}^4f_{5}\right)\;\;,
\eqa
where
\bq
f = f(x)\;\;\;,\;\;\;f_n = {1\over f(x)}f^{(n)}(x)\;\;.
\eq
Note the `homogeneity' in the number of derivatives in each term: this also
follows from simple dimensional arguments. The highest derivative
occurring in $\phi_L$ is $f_{2L-1}$.\\

The requirement of one-loop nullification, $\phi_1=0$, gives immediately that
$f'(x)=0$, so that the free action is the only possibility, as we know.
Let us therefore study nullification at two loops, that is, $\phi_2=0$.
This implies the following differential equation for $g(x)=f_1$:
\bq
3g'' - 5gg' + g^3 = 0\;\;.
\eq

Writing $g'=g^2s(g)$ we can rewrite this as
\bq
3gss' + 6s^2-5s+1 = 0\;\;.
\eq
Two obvious solutions are $s=1/2$ and $s=1/3$: the corresponding actions
are $S_2(\f)$ and $S_3(\f)$, discused above. Otherwise,
we can integrate the equation to get
\bq
\log(g) = \log(3s-1) - {3\over2}\log(2s-1) + c\;\;,
\eq
where $c$ is the constant of integration. This in turn tells us that
$s$ is the solution of a third-order algebraic equation involving $g$ and $c$,
so that there are in principle 3 solutions for $s(g)=-(1/g)'$. Working back to
$\phi_0(x)$ we pick up 2 additional constants of integration, which 
correspond to a trivial scaling transform $\phi_0(x)\to a_1\phi_0(a2x)$.
Disregarding this, we conclude that there are 3 different one-parameter
classes of actions that show two-loop nullification. Note that these
are not recursive, since any recursive action implies $s=$constant.

If, in addition to two-loop nullification, there is also nullification
at some higher $L$, we may employ $\phi_2=0$ to express
$f_{3,4,5,\dots}$ in terms of $f_1$ and $f_2$. Nullification at $L$ loops
therefore implies, because of the homogeneity, that
\bq
\phi_L=0 \Rightarrow P\left({f_2\over f_1^2}\right)=0\;\;,
\eq
where $P$ is some finite polynomial. This in turn means that $g'/g^2$ is
a constant, so that the action is necessarily recursive, and $S_{2,3}(\f)$
appear as the only possibilities. We conclude that {\em if the $L$-loop
amplitudes vanish for $L=2$ and one higher value of $L$, all
loop corrections are identically zero beyond the one-loop level}.\\

For the actions $S_{0,1,2,3}(\f)$
we see that both $\phi$ and $\phi\opi$ have interesting
nullification patterns. It is natural to wonder whether there are
other theories in which the effective action has no corrections beyond
the one-loop level, that is: is there a theory in which
\bq
\Gamma'(\phi) = S'(\phi) + \hbar\Gamma_1'(\phi) = S'(\phi_0)\;\;.
\eq
This question can be answered by inserting the loop expansion
for $\phi$ involving $f(x)$ and its derivatives, and making a Taylor
expansion in $\hbar$ in the above equation. We immediately find, from the
term linear in $\hbar$:
\bq
\Gamma_1'(\phi_0) = -{f'(x)\over2f(x)^2}\;\;,
\eq
so that not only the derivatives of $S$ but also those of $\Gamma_1$ are
completely expressed in terms of $f$ and $f_j$. The $\hbar^2$ term then
results in
\bq
3g''(x) = -4g(x)^3 + 11g(x)g'(x)\;\;\;,\;\;\;g(x)=f'(x)/f(x)\;\;,
\eq
so that all terms $\hbar^L\;,\;L\ge3$ are completely expressed
in terms of $g'/g^2$: again we are naturally led to recursive actions.
From the $\hbar^3$ and $\hbar^4$ terms we find the conditions
\bqa
0 &=& -2g^5(6v+1)(2v-1)\;\;,\nl
0 &=& -8g^7(2v-1)(1395v^2+690v-646))\;\;,
\eqa
where $g'=vg^2$ so that $v=1/u$. The only common solutions are 
$g=0$, corresponding to the free action $S_0$, and $u=2$, corresponding
to the action $S_2$, and we know that for these theories the
effective action indeed 
stops at one loop. Thus we have proven that these are, in fact,
the only theories with this property.

\section{Nullification for recursive actions}
Computing the higher-order amplitudes from Eq(\ref{biggy}) is in principle
straightforward, but for large $L$ it becomes impractical. The number
of terms in the expression (\ref{biggy}) for given $L$ is easily seen
to be equal to the coefficient of $x^L$ in the function
$$
\prod\limits_{m\ge1}\left({1\over1-x^m}\right)^{m+1}\;\;,
$$
and hence grows much 
faster than the number of partitions of $m$ which is known to
grow as $\sim\exp(\pi\sqrt{2n/3})$. For recursive actions, however,
we can simplify the treatment, as follows. The SD equation, written in terms
of $\phi(x)$, reads
\bq
S'\left(\phi(x)+\hbar\ddv{x}\right)e(x) = x\;\;,
\eq
where $e(x)=1$ is the unit function. This means that the SD equation
is built up from 
\bq
R_n = \left(\phi(x)+\hbar\ddv{x}\right)^ne(x)\;\;\;,\;\;\;n=1,2,3,\ldots
\eq
and we must evaluate these objects efficiently. As before, we put $\mu=p=1$.
Now, the $R_n$ are all of the form
\bqa
R_{n+1} & = & g_{0,n\al}\suml_{L\ge0}\hbar^Lg_{0,L\be}P_{n,L}(z)\;\;,\nl
g_{m,n} & \equiv & \phi_0(x)^m{1\over(1+x)^n}\;\;,\nl
z & \equiv & g_{1,-\al}\;\;.
\eqa
This hinges on the fact that
\bq
\ddv{x}g_{m,n} = (m-nz)g_{m-1,n+1+\al}\;\;\;,\;\;\;
\ddv{x}z = (1+\al z)g_{0,1}\;\;.
\eq
By inspection of the form of $R_2$ we obtain the following
recursive definition\footnote{The logical step function $\theta(A)$ is
one if $A$ is true, else zero.} of the polynomials $P_{n,L}$:
\bqa
P_{n,L}(z) &=& \theta(n=1)\left\{\theta(L=0)z
+ \theta(L\ge1)t_L\right\}\nl
& + & \theta(n\ge2)\left\{\suml_{M=0}^LP_{1,M}(z)P_{n-1,L-M}(z)\right.\nl
& + & \theta(L\ge1)((1-\al)(1-L)-(n-1)\al)P_{n-1,L-1}(z)\nl
&+& \left.\theta(L\ge1)(1+\al z)\ddv{x}P_{n-1,L-1}(z)\right\}\;\;.
\eqa
Here we have written $t_L = t_L(u)$ for $\phi_L(0)$, with $u=1+\al$.
Using this recursion, we can compute the $P_{n,L}(z)$ to quite high order
in $L$: notice that for given $L$, they have to be computed up to $n=2L$.
Since the SD equation holds for any $x$ we may evaluate it at $x=0$, 
where $z=0$ and $g_{0,k}=1$ for any $k$: it then becomes
\bq
\suml_{n\ge2}{\lambda_{n+1}\over n!}\left\lfloor R_{n}\right\rfloor_{x=0}
= 0\;\;.
\eq 
This allows us to successively determine the $t_L$. 
We have implemented this approach in a {\tt FORM} program \cite{form}.
Note that the complexity of the algorithm in its most straightforward
form is of order $L^6$; by various optimizations we managed to go up to
$L=50$ which takes about 24 hours of {\tt FORM}. Note that
for $L=50$ Eqn(\ref{biggy}) contains 213,927,397,257 terms! 

The lowest-order polynomials $t_L(u)$ read
\bqa
t_1(u) &=& -{u\over2}\;\;,\nl
t_2(u) &=& -{u\over24}(u-2)(u-3)\;\;,\nl
t_3(u) &=& -{u\over24}(u-1)(u-2)^2(u-3)\;\;,\nl
t_4(u) &=& -{u\over1920}(u-2)(u-3)(7u^2-21u+12)(23u^2-69u+50)\;\;,\nl
t_5(u) &=& -{u\over1440}(u-1)(u-2)^2(u-3)
\nl &&(367u^4-2202u^3+4685u^2-4146u+1260)\;\;,\nl
t_6(u) &=& -{u\over580608}(u-2)(u-3)
\nl &&(601285u^8-7215420u^7+37068226u^6-106328304u^5
\nl &&+185954749u^4-202661124u^3+134127612u^2
\nl &&-49166352u+7620480)\;\;,\nl
t_7(u) &=& -{u\over60480}(u-1)(u-2)^2(u-3)
\nl &&(318344u^8-3820128u^7+19590653u^6-55981845u^5
\nl &&+97298966u^4-105068217u^3+68637897u^2
\nl &&-24716070u+3742200)\;\;,\nl
t_8(u) &=& -{u\over199065600}(u-2)(u-3)
\nl &&(6389072441u^{12}-115003303938u^{11}+935605664709u^{10}
\nl &&-4546312395750u^9+14686419780735u^8-33204111807078u^7
\nl &&+53832598760431u^6-63007849676250u^5+52791473853204u^4
\nl &&-30847914995544u^3+11919566344320u^2
\nl &&-2731077648000u+280215936000)\;\;,
\eqa
and so on. 

From duality we have seen that $t_L(u)$ is of the form
\bq
t_L(u) = u(u-2)(u-3)Q_L(u)\;\;,
\eq
with $Q_L(u)$ a polynomial of degree $2L-4$ in $u$, which is symmetric
under $u\leftrightarrow3-u$: it is therefore a polynomial of degree
$L-2$ in the variable $u(3-u)$, which makes finding the roots simpler.
We therefore discuss only the `lower half' of the roots of $Q_L(u)$.
Surprisingly, all roots are real up to $L=18$, where a pair of conjugate
roots appear. Other pairs appear at $L=27,34,41$ and 48.
In figure 1 we give the distribution of the real values of the roots of $Q_L(u)$
that are smaller than 3/2. It is suggestive to follow roots over
trajectories as $L$ increases. At $L=18$ the second and third lowest lines
appear to merge, leading to conjugate complex roots. For higher $L$,
the fourth line does not
merge with these two, but actually crosses them (this is borne out by
inspecting which roots are real, and which ones complex). For $L=27$ a
similar phenomenon occurs, and so on. For large $L$, there is an apparent
asymptotic upper limit 3/2 which is just an artefact of our restriction to
Re$(u) <3/2$. There is also a lower asymptotic bound 1/2. 
This bound can, in fact, be understood: it is easy to see that,
for $0<u<1/2$, the single vertex $-\lambda_{n+1}$ occurring in
a diagram has {\em the same sign} as the product $\lambda_n\lambda_3$,
corresponding to `opening up' the vertex by insertion of a propagator.
This imples that in that case all diagrams contributing to a given
amplitude have precisely the same sign, so no cancellation is possible
and no root of $Q_L(u)$ can be in $(0,1/2)$ for any $L$: by duality, the
same holds for the interval $(5/2,3)$. On the other hand, it is not clear
why there should be no roots with negative real part 
(or real part larger than 3), or why there should not be any complex roots
with real part between 0 and 1/2.
Figure 2 shows the distribution of the imaginary parts of the roots, where
the branching structure becomes especially apparent.
\begin{center}
\begin{figure}
\plaat{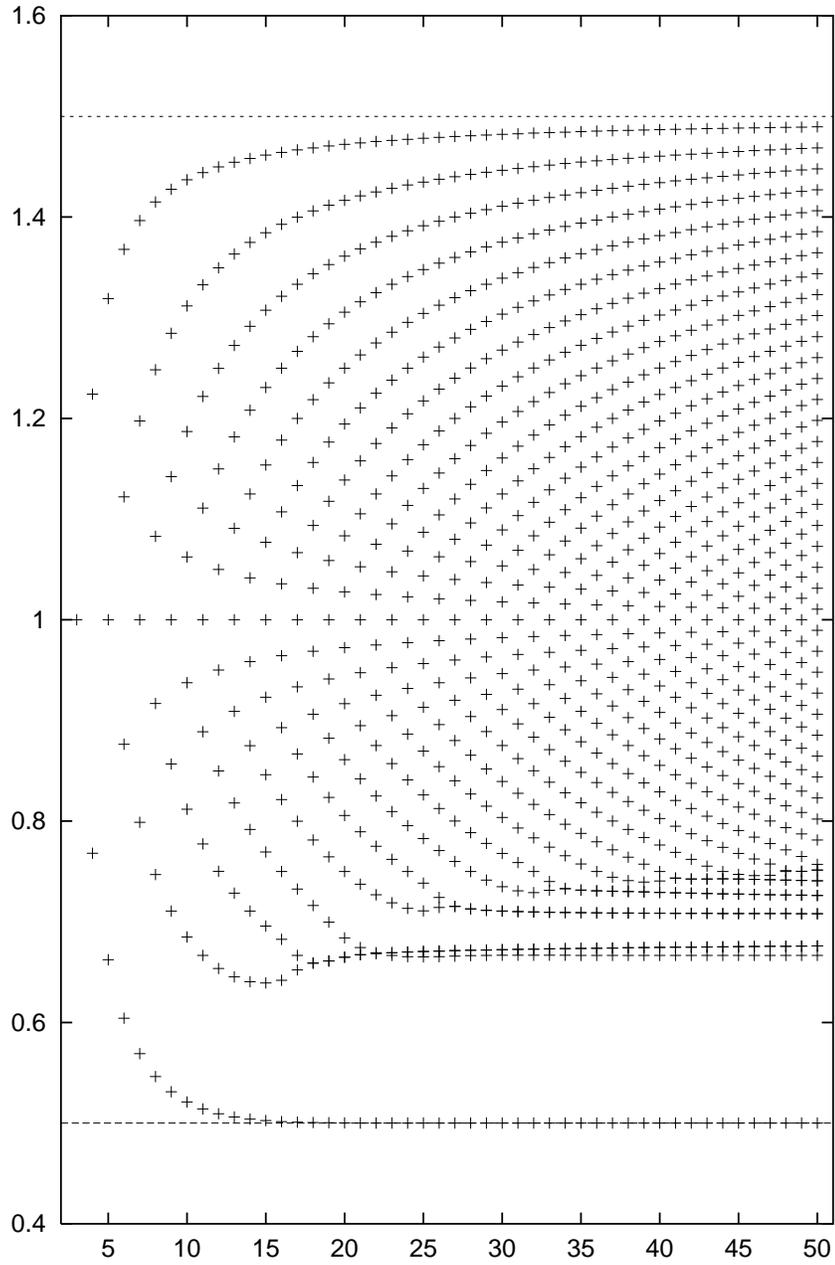}{12}
\caption[.]{Distribution of the real values of $u$
for which $Q_L(u)$ vanishes, for $3\le L\le50$. Horizontal: $L$, vertical:
Re$(u)$.}
\end{figure}
\end{center}
\begin{center}
\begin{figure}
\plaat{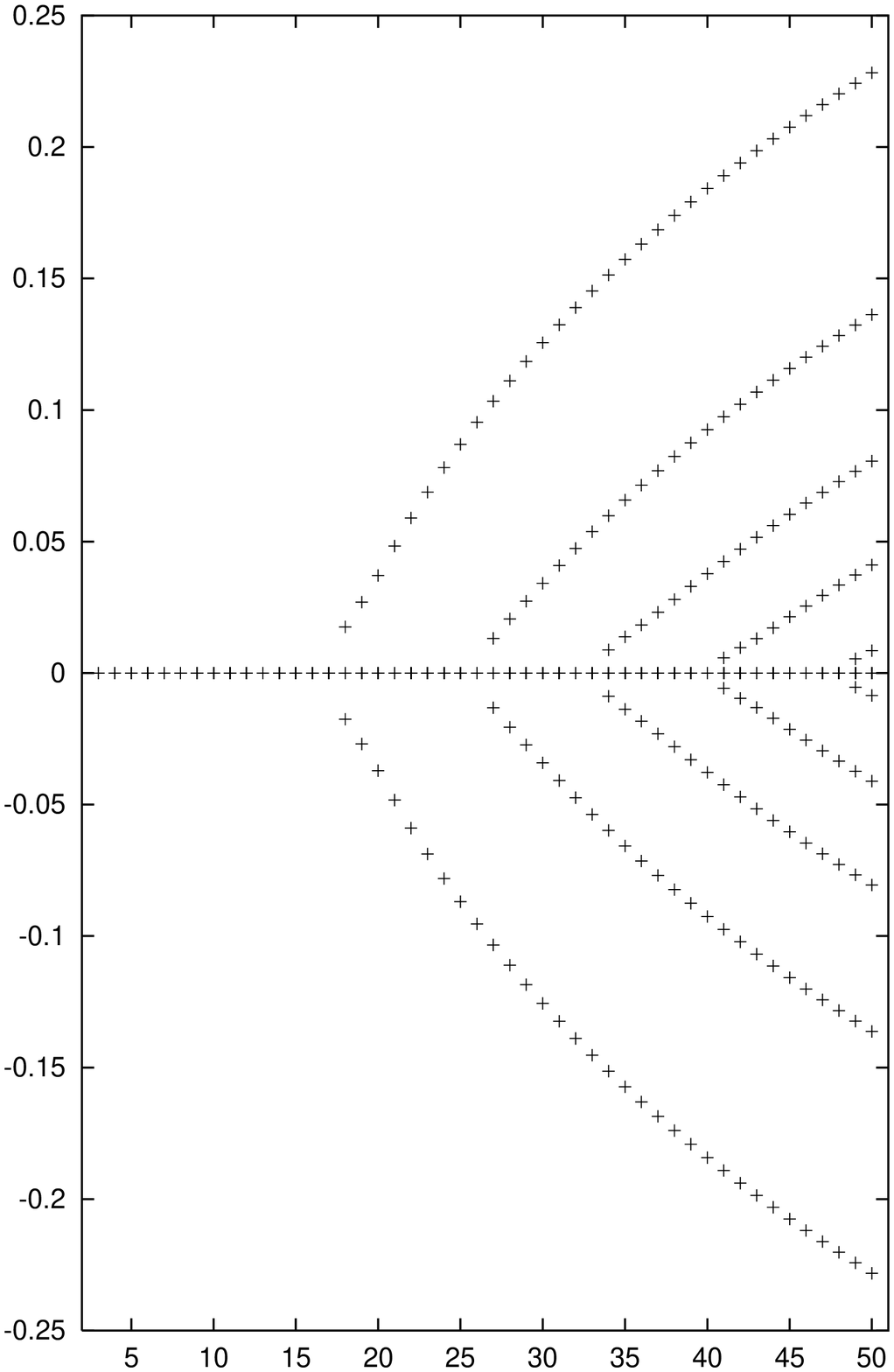}{12}
\caption[.]{Distribution of the imaginary values of $u$
for which $Q_L(u)$ vanishes, for $3\le L\le50$. Horizontal: $L$, vertical:
Im$(u)$.}
\end{figure}
\end{center}

\section{Renormalization}
Because of the fairly simple structure of $\phi(x)$ in zero dimensions,
it is often possible to carry through a renormalization program for
recursive actions, as we shall now show.\\

Reinserting generic values for $\al$ and $\mu$, the action $S_2(\f)$
has for its solution
\bq
\phi(x) = {x-\al\hbar\over\mu+\al x}\;\;.
\eq
As a first step, we add a tadpole counterterm to the action, which has the
effect of shifting the variable $x$ by a constant. The tadpole
renormalization condition is that $\phi(x)$ should have no tadpole left after
renormalization. Denoting the renormalized generating function by $\phi_r(x)$,
we therefore have
\bq
\phi_r(x) \equiv \phi(x+c)\;\;\;,\;\;\;
\phi_r(0)\equiv0\;\;\Rightarrow\;\;c=\al\hbar\;\;,
\eq
so that
\bq
\phi_r(x) = {x\over\mu+\al^2\hbar+\al x}\;\;.
\eq
The second renormalization condition is that of mass renormalization.
Denoting the renormlized, physical mass by $m$, we therefore require
\bq
\phi_r'(0) = {1\over m^2}\;\;,
\eq
which fixes $\mu$:
\bq
\mu = m^2 - \al^2\hbar\;\;.
\eq
The resulting renormalized generating function and the renormalized effective
action can therefore be written as
\bq
\phi_r(x) = {x\over m^2+\al x}\;\;\;,\;\;\;
\Gamma(\f) = {m^2\over\al^2}\left(-\al\f-\log(1-\al\f)\right)\;\;.
\eq
Hence, {\em all\/} loop corrections have been completely absorbed.
As mentioned before, a similar finding occurs for $d=0$,$u=4/3$.\\

The action $S_3(\f)$ also leads to fairly simple results. We have
\bq
\phi(x) = {1\over\al}-{4\mu^2\over\al}{1\over(2\mu+\al x)^2}
 -{3\al\hbar\over2}{1\over2\mu+\al x}\;\;.
\eq
Writing the tadpole counterterm as $(c-2)\mu/\al$ we have for the renormalized
$\phi$:
\bq
\phi_r(x) = {1\over\al}-{4\mu^2\over\al}{1\over(c\mu+\al x)^2}
-{3vm^2\over2\al}{1\over c\mu+\al x}\;\;,
\eq
where the renormalization conditions are, as before,
\bq
\phi_r(0)=0\;\;\;,\;\;\;\phi_r'(0)={1\over m^2}\;\;,
\eq
and we have introduced the dimensionless parameter $v=\al^2\hbar/m^2$.
The two renormalization conditions imply two coupled equations for $c$ and $\mu$:
\bqa
0 &=& -2\mu^2c^3+3vm^4c+16\mu m^2\;\;,\nl
0 &=& 16\mu m^2+(-8\mu^2+3vm^4)c-3\mu c^2vm^2\;\;.
\eqa
These can be combined to give $c$ as a function of $\mu$, and a 
quadratic equation for $\mu^2$:
\bqa
c &=& {8\mu m^2(4-3v)\over16\mu^2-6vm^4+9v^2m^4}\;\;,\nl
0 &=& -128\mu^4-192m^4v\mu^2+128m^4\mu^2+18v^2m^8-27v^3m^8\;\;.
\eqa
The perturbative solution for $\mu$ is
\bq
\mu = {m^2\over4}\sqrt{8-12v+\sqrt{64-192v+180v^2-54v^3}}\;\;,
\eq
so that the perturbative expansion of $\mu$ and $c$ contain an infinite number
of terms. we conclude that, although the unrenormalized amplitudes vanish
for two or more loops, renormalization reintroduces nonzero amplitudes at
all loops.\\

For the action $S_1(\f)$ we have
\bq
\phi(x) = {1\over p}\left[\psi\left({\mu+px\over\hbar p^2}\right)
+\log\left({\hbar p^2\over\mu}\right)\right]\;\;.
\eq
The conditions for the renormalized function $\phi_r(x)=\phi(x+c)$
now read
\bqa
\phi_r(0) = 0 &\Rightarrow& \psi(w)
= \log\left({\mu\over p^2\hbar}\right)\;\;,\nl
\phi_r'(0) = {1\over m^2} &\Rightarrow& \psi'(w) = {p^2\hbar\over m^2}\;\;,
\eqa
with $w=(\mu+pc)/(p^2\hbar)$. Since $\psi'$ is monotonic for $w\ge-1$ and
takes all real values, the second equation gives $w$ uniquely for given $m$;
and the first one then gives $\mu$, and hence also $c$. 
Following through this program in perturbation theory gives the following
interesting result. Using the dimensionless number $v=p^2\hbar/m^2$
and the asymptotic expansions for $\psi$ and $\psi'(w)$, we obtain the
following results:
\bqa
w &=& {1\over v} + {1\over2} - {1\over12}v + {11\over720}v^3
 - {379\over30240}v^5 + {24369\over1209600}v^7  \ldots\;\;,\nl
{\mu\over m^2} &=& 1 - {1\over24}v^2 + {71\over5760}v^4 
- {31741\over2903040}v^6 + {25265783\over1393459200}v^8 + \ldots\;\;,\nl
c{p\over m^2} &=& {1\over2}v - {1\over24}v^2 + {17\over5760}v^4
- {4643\over2903040}v^6 + {559157\over278691840}v^8 + \ldots\;\;,
\eqa
and, putting $m=p=1$ for simplicity, we find for the renormalized
generating function
\bqa
\phi_r(x) &=& \log(1+x) + {x^2v^2\over24}{1\over(1+x)^2}\nl
&& -{x^2v^4\over2880}{76+88x+33x^2\over(1+x)^4}\nl
&& +{x^2v^6\over362880}{3790x^4+18192x^3+34572x^2+31636x+12861\over(1+x)^6}
\nl &&+ \cdots
\eqa
The higher powers of $v$ in the results for $(vw)$,$\mu$,$c$ and $\phi$ 
all appear to be even. We have checked this up through order $v^{30}$. The
conclusion is that for the action $S_1$, {\em all\/} odd-loop corrections
vanish after tadpole and mass renormalization, thereby even improving on
the unrenormalized pattern. This is in accordance with our
conjecture that for this
theory the only odd-loop contribution to the effective action is the
one-loop tadpole, which is removed by renormalization. 

\section{Conclusions}
We have identified recursive theories,
an essentially one-parameter
class of self-interacting scalar theories in which
zero-momentum amplitudes are related in a simple and systematic manner.
In the case of zero-dimensional theories, several of these theories can be
solved exactly and display an interesting pattern of vanishing higher-loop
amplitudes and 1PI amplitudes. We have identified a duality property in which 
composite objects in a given theory are the elementary fields in its dual.
A study of the dependence of the higher-loop amplitudes
on the parameter, $u$, of the theory reveals a remarkable pattern of
roots.

\end{document}